\documentstyle[aps, prl, epsf, multicol]{revtex}

\begin{document}

\title{Conductance fluctuations in a quantum dot under almost periodic ac
pumping.} 
\author{Xiao-Bing Wang$^1$ and V. E. Kravtsov$^{1,2}$}
\address{$^1$ Tha Abdus Salam International Centre for Theoretical Physics,
P.O. Box 586, 34100 Trieste, Italy\\
$^2$ Landau Institute for Theoretical Physics, 2 Kosygina Street, 117940
Moscow, Russia}
\date{\today}
\maketitle

\begin{abstract}
It is shown that the variance of the linear dc conductance fluctuations in
an open
quantum dot under a high-frequency ac pumping depends significantly
on the spectral content of the ac field. For a sufficiently strong ac
field $\gamma\tau_{\varphi}\ll 1$, where $1/\tau_{\varphi}$ is the
dephasing rate
induced by ac
noise and $\gamma$ is the electron escape rate, the dc conductance
fluctuations are much stronger for the harmonic pumping than in the case
of the noise ac field of the same intensity. 
The reduction
factor $r$  in a  static magnetic field  takes the universal value of 
2
only for the white--noise pumping. For the strictly
harmonic pumping $A(t)=A_{0}\cos\omega t$ of sufficiently large intensity
the
variance is almost
insensitive to the static magnetic field $r-1=
2\sqrt{\tau_{\varphi}\gamma}\ll 1$.
For the quasi-periodic
ac field of the form
$A(t)=A_{0}\,[\cos(\omega_{1} t)+\cos(\omega_{2} t)]$ with
$\omega_{1,2}\gg \gamma$ and $\gamma\tau_{\varphi}\ll 1$ 
we predict the novel effect of enchancement of conductance fluctuations
at commensurate frequencies $\omega_{2}/\omega_{1}=P/Q$.

\noindent PACS numbers: 72.15.-v, 72.30.+q, 73.23.-b

\end{abstract}

\begin{multicols}{2}
\narrowtext

One of the most important discoveries in mesoscopic physics is the universal
conductance fluctuations\cite{web,alts1}. These
fluctuations of the order of $\delta G\sim e^2/\hbar$, are due to
stochastic quantum
interference. They depend only on the effective
dimensionality and  general symmetries, but not on the microscopic
details of the system. The fluctuations can manifest themselves as 
reproducible
aperiodic magneto-resistance patterns(magnetic fingerprints). 
Variation
of the fluctuations with magnetic fields and chemical potential, temperature 
{\it etc}.
were calculated by the diagrammatic technique\cite{alts2}
and observed in many disordered
electronic
systems\cite{gior}. In the presence of a strong enough static magnetic
field, the time
reversal symmetry
is broken and the variance  $\langle \delta
G^2\rangle$ ($\langle...\rangle$ stands for the disorder
average)  of conductance fluctuations is
reduced by the factor $r=2$. 

Recently there has been a considerable interest in non-equilibrium
mesoscopics. The effect of adiabatic charge pumping \cite{Thou} has been
experimentally observed \cite{Marc} and analyzed theoretically
\cite{Brou,SAA}.
Weak localization under ac pumping \cite{VA}  and the
photovoltaic effect \cite{VAA} in a quantum dot have been
theoretically studied. 
The non-equilibrium noise has been suggested \cite{krav3} as a cause of
both
the low
temperature dephasing saturation \cite{moh1} and the anomalously large
ensemble averaged persistent current \cite{Levi}. 

Here we study the effect of the high-frequency ac field on the mesoscopic
fluctuations of {\it linear} dc conductance when both the {\it weak dc
voltage} and a {\it strong enough high-fequency pump field} are
applied to the
open quantum dot \cite{Falko}. 
We will study the dependence of the variance of the dc conductance
fluctuations on the ac field intensity for  the noise-like and an almost
periodic ac field.
In particular we focus on the reduction factor 
$r=1+{\cal C}/{\cal D}$ for the variance of conductance
fluctuations after 
turning on a strong time-reversal breaking (e.g. the static magnetic
field) that kills the cooperon contribution to the variance $\langle
\delta G^2 \rangle_{C}\equiv{\cal C}$
while
leaving the diffuson one $\langle\delta G^2 \rangle_{D}\equiv {\cal D}$
unchanged. 
\begin{figure}[tbp]
\centerline{\epsfysize=4.0cm \epsffile{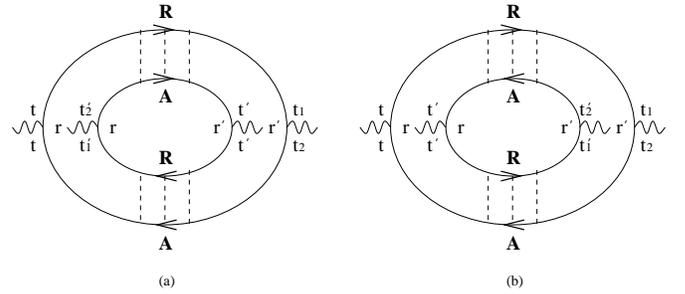}}
\caption{The cooperon (a) and the diffuson (b) contributions to the
variance
of conductance fluctuations.}
\end{figure}
The
Landauer conductance $g=(\gamma V/4)^2 \,[K({\bf r},{\bf r'})+K({\bf
r'},{\bf r})]$ of a dot
of the volume $V$ with
small contacts at ${\bf r}$ and ${\bf
r'}$ and the electron escape rate $\gamma$, can be expressed in terms of
the exact retarded and advanced electron Green's
functions $G^{R,A}({\bf r},{\bf r'};t,t')$ in the time domain \cite{VA,YKK}:
\begin{eqnarray}
\label{cond}
K({\bf r},{\bf r'})=\int
dt_{1}dt_{2}\,\overline{G^{R}({\bf
r},{\bf
r'};t,t_{1})G^{A}({\bf
r'},{\bf r};t_{2},t)}\,F_{t_{1}-t_{2}}
\end{eqnarray}
where $F_{t}=\pi
Tt\sinh^{-1}(\pi T t)$ is the
Fourier-transform of
the derivative of the Fermi distribution function for electrons in the leads and
$\overline{f(t)}=\int_{-{\cal T}/2}^{{\cal T}/2}\frac{dt}{\cal T} f(t)$
denotes time averaging
during  the observation time ${\cal T}\rightarrow\infty$.

We consider a chaotic or disordered quantum dot with the number of open
channels $M\gg 1$ 
and the electron
escape rate $\Delta \ll \gamma=\Delta M \ll E_{c}$  
where $\Delta$ is the mean level separation 
and the Thouless energy $E_{c}$ is the inverse ergodic time.
In this situation the charging effects are negligible, the perturbative 
diagrammatic analysis (see Fig.1) is possible and the ergodic zero-dimensional
approximation applies. We also assume
the dephasing rate $\gamma_{int}(T)$ caused by electron interaction to be
smaller than
the escape rate $\gamma$.   
At such conditions in the absence of ac pumping one obtains 
 ${\cal C}={\cal D}\sim
(\gamma/\Delta)^2$ at zero temperature. Thus the reduction
factor $r=2$. 

However, the results may change if the dot is subject to a 
time-dependent perturbation. 
In that case the Green's functions 
$G^{R,A}({\bf r},{\bf r'};t,t')$ are inhomogeneous in time, so that the
cooperon 
$$\langle G^{R}({\bf r},{\bf 
r'};t_{+},t_{+}^{\prime})
G^{A}({\bf r},{\bf r'};t_{-},t_{-}^{\prime})
\rangle =\frac{1}{2}\, \delta(t-t')\, C_{t}(\eta, \eta ^{\prime })$$
and the diffuson
$$\langle G^{R}({\bf r},{\bf
r'};t_{+},t^{\prime}_{+})
G^{A}({\bf r'},{\bf r};t_{-}^{\prime},t_{-})
\rangle=\delta(\eta-\eta')\,D_{\eta}(t,t^{\prime })$$ 
are no longer functions of the difference of two times
but in general depend on four time variables $t_{\pm}=t\pm\eta/2,
t_{\pm}^{\prime}=t^{\prime}\pm\eta^{\prime}/2$. They can be found from
\cite{{alts5}}: 
\begin{eqnarray}\label{coop}
\left\{2 \frac \partial {\partial \eta }+\gamma+ D
\left[i\nabla +  A\left(t+\frac \eta 2\right)+
A\left(t-\frac \eta
2\right)\right]^2\right\}\\ \nonumber\times
C_{t}(\eta, \eta ^{\prime };{\bf r},{\bf r'})=2 \delta (\eta -\eta
^{\prime
})\delta({\bf r}-{\bf r'}),
\end{eqnarray}
and

\begin{eqnarray}
\label{diff}
\left\{\frac \partial {\partial
t}+\gamma+D\left[i\nabla +
A\left(t+
\frac \eta 2\right)-A\left(t-\frac \eta 2\right)\right]
^2\right\} \\ \nonumber\times
D_{\eta}(t,t^{\prime };{\bf r},{\bf r'})= \delta (t-t^{\prime
})\delta({\bf r}-{\bf r'}).
\end{eqnarray}
For concreteness we consider the `dot'
in a form of a quasi-1d ring of the circumference $L$ with two leads at
points ${\bf r}$ and
${\bf r'}$. 
Qualitatively, the results do not change if the dot has the
form of a disc or the ac pumping is produced by applying the gate voltage
that changes the dot's shape \cite{AAV}.
\begin{figure}[tbp]
\centerline{\epsfysize=2.5cm \epsffile{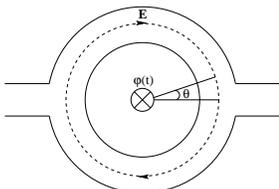}}
\caption{The circular `dot' under ac pumping: the circular electric field
results from the oscillating magnetic flux $\varphi(t)$.}
\end{figure}
The time-dependent magnetic flux $\phi(t)=
A(t) L $ threads the ring
and results in the circular electric field $E_{\theta}(t)= -d/dt A(t)$
which does
not depend on the position $x=L\theta/2\pi$ (see Fig.2).
The periodic boundary condition
$D_{\eta}(t,t';x-x')=D_{\eta}(t,t';x+L-x')$ and
$C_{t}(\eta,\eta';x-x')=C_{t}(\eta,\eta';x+L-x')$ allows to switch in
Eqs.(\ref{coop},\ref{diff}) to the
Fourier components 
$D_{\eta}(t,t^{\prime };q)$ and $C_{t}(\eta, \eta ^{\prime };q)$ 
with $q_{n}=(2\pi/L)\,n$, $(n=0,\pm 1, \pm 2,...)$ and the condition
$E_{c}/\gamma \gg 1$ 
allows to leave only the
zero-mode components $D_{\eta}(t,t^{\prime })=D_{\eta}(t,t^{\prime };q=0)$
and $C_{t}(\eta, \eta ^{\prime })=C_{t}(\eta, \eta ^{\prime };q=0)$ (ergodic
approximation):
\begin{equation}
\label{dif1}
D_{\eta}(t,t^{\prime
})=\Theta_{\eta-\eta'}\,e^{-\int_{t'}^{t} \Gamma_{d}(\eta,\xi)\,d\xi}
\end{equation}
\begin{equation}
\label{coop1}
C_{t}(\eta,\eta^{\prime
})=\Theta_{t-t'}\,e^{-\frac{1}{2}\int_{\eta'}^{\eta}
\Gamma_{c}(t,\xi)\,d\xi},
\end{equation}
where $\Theta_{t}$ is the step-function and \cite{AAV}:
\begin{equation}
\label{gamd}
\Gamma_{d}(\eta,\xi)=\gamma+D\left(A(\xi+\eta/2)-A(\xi-\eta/2)\right)^2,
\end{equation}
\begin{equation}
\label{gamc}  
\Gamma_{c}(t,\xi)=\gamma+D\left(A(t+\xi/2)+A(t-\xi/2)
\right)^2.
\end{equation}
Then one obtains from the diagrams of Fig.1:
\begin{equation}\label{d2}
{\cal D}\propto\int dt^{\prime }dt^{\prime \prime }
\overline{D_{\eta}
(t,t^{\prime
})D_{-\eta
}(t,t^{\prime \prime })}\;F^{2}_{t^{\prime }-t^{\prime \prime}}. 
\end{equation}
\begin{eqnarray}\label{c2}
{\cal C}\propto\int dt^{\prime}d\eta ^{\prime }
\overline{C_{t^{\prime}+t}
(-\eta^{\prime}-2t',-\eta+2t')C_{t}(\eta,\eta
^{\prime})}\,F^{2}_{2t^{\prime}}.
\end{eqnarray}
where averaging over $t$ and $\eta$ is assumed.

We observe that while the variables in the arguments of diffusons are
decoupled, they are strongly coupled in the argument of a cooperon.
As usual this leads to the restricted region of integration in
Eq.(\ref{c2}) and to a general inequality ${\cal C}/{\cal D} <1$ or
$r< 2$.

However, there is a dramatic difference between the noise-like ac
field with 
the short correlation time $\tau_{0}\sim \omega^{-1}\ll \tau_{\varphi}$
and 
the harmonic ac field $A(t)=A_{0}\cos(\omega t)$ with $\omega\gg
\gamma$. At such a high-frequency pumping one can replace
$\Gamma_{d}(\eta,\xi)$ and $\Gamma_{c}(t,\xi)$ in
Eqs.(\ref{dif1},\ref{coop1}) by the 
time averages
$\Gamma_{d}(\eta)=\overline{\Gamma_{d}(\eta,\xi)}$ and 
$\Gamma_{c}(t)=\overline{\Gamma_{c}(t,\xi)}$. This approximation will be
referred to as the {\it high frequency ansatz}.
In the case of the white--noise pumping the average of the cross-terms
$\overline{A_{\xi+\eta/2}A_{\xi-\eta/2}}$ and
$\overline{A_{t+\xi/2}A_{t-\xi/2}}$ 
in Eqs.(\ref{gamd},\ref{gamc}) is zero and we obtain {\it the same,
time-independent} decay
rates 
$\Gamma_{d}=\Gamma_{c}=\gamma+1/2\tau_{\varphi}$
for the cooperons and the diffusons with $\eta\neq 0$,
where
\begin{equation}
\label{phi}
\tau_{\varphi}^{-1}=4 D \overline{A^2}\propto I
\end{equation}
is proportional to the pumping intensity $I$.
 As the result the
reduction factor $r=2$ is unchanged.
In contrast to that, for the harmonic pumping
the decay rates
$\Gamma_{c}(t)=\gamma+\tau_{\varphi}^{-1}\,\cos^2 (\omega t)$ and
$\Gamma_{d}(\eta)=\gamma+\tau_{\varphi}^{-1}\,\sin^2 (\omega \eta/2)$ are
{\it
periodic} functions of the `additional' time variables $t$ and $\eta$.
Thus the pumping-induced dephasing is effectively {\it switched off} 
inside the `no-dephasing' time intervals where $\Gamma_{d,c}(t)\ll \gamma$.
At strong pumping
$\tau_{\varphi}^{-1}\gg \gamma$ these time intervals make the
main
contribution to mesoscopic fluctuations.

Below we present
analytical results for the reduction factor $r$ in the high-frequency
limit $\omega\gg
\tau_{\varphi}^{-1}\gg \gamma$\cite{AAV}.

Using the high-frequency ansatz we obtain from Eqs.(\ref{d2},\ref{c2}):
\begin{eqnarray}
\label{LowTd}
\frac{{\cal 
D}}{g_{0}^2}=\int_{-{\cal
T}/2}^{{\cal T}/2}
\frac{d\eta}{2{\cal
T}}\,\left(\frac{\gamma^2}{\Gamma_{d}(\eta)} \right)
\int_{0}^{\infty}e^{-t\,\Gamma_{d}(\eta)}\,F^{2}_{t}\,dt.
\end{eqnarray}
\begin{eqnarray}
\label{LowTc}
\frac{{\cal C}}{g_{0}^2}=
\int_{-{\cal
T}/2}^{{\cal T}/2}\frac{dt}{{\cal T}}\int_{-\infty}^{t}dt'\,
\frac{\gamma^2 F^{2}_{2t-2t'}\,e^{-2(t-t')\,\Gamma_{c}(t')}}
{\frac{1}{2}(\Gamma_{c}(t)+\Gamma_{c}(t'))},
\end{eqnarray}
where $g_{0}=\pi\gamma/4\Delta$ is the mean conductance,
${\cal T}\rightarrow\infty$ is an observation time, and $F_{t}=\pi T
t \sinh^{-1}(\pi T t)$.

Eqs.(\ref{LowTd},\ref{LowTc}) can be simplified in the limit of low
temperatures $T\ll\gamma$ where $F_{t}\approx 1$ and in the limit of high
temperatures $T\gg\omega(\gamma\tau_{\varphi})^{-1/2}$ where
$F^2_{t}\approx (\pi/6T)\, 
\delta(t)$ and we have:
\begin{equation}
\label{highT}
\frac{\langle \delta G^2\rangle_{D,C}}{g_{0}^2}=\frac{\pi \gamma^2}{12 T}
\int_{-{\cal T}/2}^{{\cal T}/2}\frac{dt}{{\cal
T}}\,\frac{1}{\Gamma_{d,c}(t)}.
\end{equation}

As has been already mentioned, in the case of the
nearly white-noise pumping 
$\Gamma_{d}=\Gamma_{c}=\gamma+1/2\tau_{\varphi}$ are
independent of $\eta$ and $t$. For relatively small
pumping
intensity $I\propto (\gamma\tau_{\varphi})^{-1}\ll \gamma/\Delta$
the
variance of
conductance fluctuations is still given by diagrams of Fig.1 and we have:
\begin{equation}
\label{noi}
\frac{{\cal D}}{g_{0}^2}=\frac{{\cal
C}}{g_{0}^2}=\left\{\matrix{\frac{1}{2}\,
\left(1+\frac{1}{2\gamma\tau_{\varphi}}
\right)^{-2}, & T\ll \gamma +\frac{1}{2\tau_{\varphi}}\cr
\frac{\pi\gamma}{12 T}\,\left(1+\frac{1}{2\gamma\tau_{\varphi}}  
\right)^{-1}, &  T\gg \gamma +\frac{1}{2\tau_{\varphi}}\cr
}\right.   
\end{equation}
For larger intensities  $(\gamma\tau_{\varphi})^{-1}\gg \gamma/\Delta$
one has to take into account dressing of vertices  \cite{mac}
and the result changes. Yet in both cases the reduction
factor $r$ is strictly 2.

For the  periodic pumping the decay rates are periodic functions of time:
$\Gamma_{d}(t)=\gamma+\tau_{\varphi}^{-1}\,\sin^2 (\omega t/2)$,
$\Gamma_{c}(t)=\gamma+\tau_{\varphi}^{-1}\,\cos^2 (\omega t)$, and
Eqs.(\ref{LowTd},\ref{LowTc}) no longer lead to the same result.
At high pumping intensities $\gamma\tau_{\varphi}\ll 1$ the integrals in
Eqs.(\ref{LowTd},\ref{LowTc}) are dominated by the vicinity of
zeros $t^{D,C}_{n}$ of $\sin^2 (\omega t/2)$ and $\cos^2(\omega t)$.
One can expand $\Gamma_{d,c}(t)$ near $t^{D,C}_{n}$,
perform integrations from $-\infty$ to $+\infty$ over $t-t^{D,C}_{n}$
and sum over $t^{D,C}_{n}$. The result depends on the relation between
the temperature of leads $T$ and $\gamma$. 

For $T\ll \gamma\ll \tau_{\varphi}^{-1}$ we obtain:
\begin{equation}
\label{dd}
\frac{{\cal D}}{g_{0}^2}\approx
\frac{1}{4}\sqrt{\gamma\tau_{\varphi}}\propto
\frac{1}{\sqrt{I}},\;\;\;\;\;\;\;\;\;\;\;\frac{{\cal
C}}{g_{0}^2}\approx \frac{1}{2}\,\gamma\tau_{\varphi}\propto
\frac{1}{I}.
\end{equation}
One can see that the cooperon contribution ${\cal C}$ is strongly
suppressed at high pumping intensities and the reduction factor $r=1+2
(\gamma\tau_{\varphi})^{1/2}$ is close to 1.

At high temperatures $T\gg \omega(\gamma\tau_{\varphi})^{-1/2}$ and strong
pumping $\gamma\ll
\tau_{\varphi}^{-1}$ Eq.(\ref{highT}) gives the same
value
\begin{equation}
\label{HT}
\frac{{\cal D}}{g_{0}^2}=\frac{{\cal C}}{g_{0}^2}=\frac{\pi\gamma}{12
T}\,\sqrt{\gamma\tau_{\varphi}}
\end{equation}
for the diffuson and the cooperon contribution to conductance
fluctuations, and $r=2$.

However, even at high temperatures the cooperon contribution can be
significantly suppressed relative to the diffuson one if the pumping field
is a sum of {\it two} harmonic parts:
\begin{equation}
\label{inc}
A(t) = A_{0}\,[\cos (\omega t) + \cos(\alpha\omega t)],\;\;\;(0<\alpha<1).
\end{equation}
In this case we have from Eqs.(\ref{gamd},\ref{gamc})
$\Gamma_{d}(t)-\gamma=
\tau_{\varphi}^{-1}[\sin^2(\omega t/2)+\sin^2 (\alpha\omega t/2)]/2$
and $\Gamma_{c}(t)-\gamma=
\tau_{\varphi}^{-1}[\cos^2(\omega t)+\cos^2 (\alpha\omega t)]/2$. In the
limit of strong pumping $\gamma\tau_{\varphi}\ll 1$ one can
express the integral in Eq.(\ref{highT}) in terms of the distribution
of complex roots $t_{n}$ of the equation $\Gamma_{d,c}(t)=\gamma$. 
This is the problem of distribution of complex roots
$z_{n}=x_{n}+iy_{n}$ with $y_{n}\sim \sqrt{\gamma\tau_{\varphi}}\ll 1$ of
the
transcendental equation:
\begin{equation}
\label{eq}
\cos z + \cos(\alpha z)\pm 2 =0,
\end{equation}
where the sign $\pm$ stands for the cooperon and the diffuson contribution
to the variance, respectively.
 
Defining the densities of complex roots $\rho_{d,c}(y)$ of Eq.(\ref{eq})
with the sign $-$ $(\rho_{d})$ or $+$ $(\rho_{c})$:
\begin{eqnarray}
\label{df}
\rho(y)=
\sum_{n}\langle\delta(x-x_{n})\delta(y-y_{n})\rangle_{x},
\end{eqnarray}
where $\langle
...\rangle_{x}=(\omega{\cal T})^{-1}\int_{|x|<\omega{\cal
T}/2}dx ...$  stands for the averaging
over $x$, we obtain:
\begin{equation}
\label{DD}
\frac{\langle \delta G^2
\rangle_{D,C}}{g_{0}^2}=\frac{\pi^2\gamma^2\tau_{\varphi}
}{3T \sqrt{1+\alpha^2}}\,\int_{-\infty}^{+\infty}\frac{
dy\, \rho_{d,c}(y)}{\left[2\gamma\tau_{\varphi}+
\frac{y^2}{4}\,(1+\alpha^2)
\right]^{1/2}}.
\end{equation}
For the case of two {\it commensurate}
frequencies
$\alpha=P/Q <1$ with $Q\sim 1$ the density $\rho(y)$ is the set of
$\delta$-functions $(2\pi Q)^{-1}\sum \delta (y-y_{n})$ separated by
gaps
$\Delta
y_{n}\sim 1/Q$.
At $\gamma\tau_{\varphi}\ll 1/Q^2$ the gap
$y_{1}\sim 1/Q$ is large compared to $(\gamma\tau_{\varphi})^{1/2}$, and
in
the leading approximation one can neglect all complex roots with
$y_{n}\neq 0$.
Equation (\ref{eq}) with the sign minus has real roots at {\it any}
rational $\alpha$.
However,
Eq.(\ref{eq}) with the sign plus
(relevant for the cooperon contribution) has real solutions
only if $P$ and $Q$ are both odd.
In this case ${\cal D}={\cal C}\propto I^{-1/2}$ as
for a strictly harmonic pumping. However the cooperon contribution is
anomalously suppressed  ${\cal C}\propto I^{-1}$ if
either $Q$ or $P$ is even.
This parity effect is also present for $T\ll \gamma$ as it is seen from 
Fig.3 obtained by numerical
integration of Eqs.(\ref{LowTd},\ref{LowTc}).
\begin{figure}[tbp]
\centerline{\epsfysize=6.0cm \epsffile{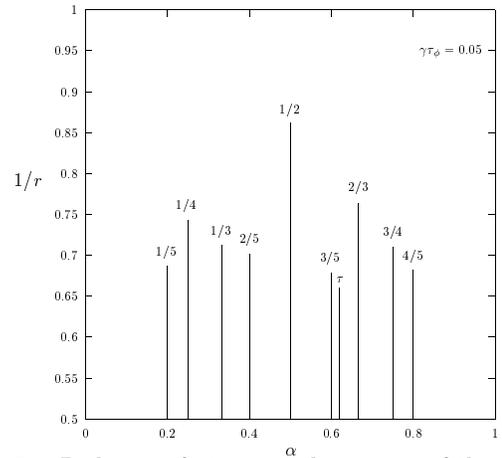}}
\caption{ Reduction factor $r$ vs the ratio $\alpha$ of
the two frequency components. Also plotted is $r$ for the golden mean
$\tau=\frac{\sqrt{5} -1}{2}$.}
\end{figure}
With increasing $Q$ the gaps $\Delta y_{n}$ shrink to zero and in the
limit $Q\rightarrow\infty$ one obtains smooth distribution
$\rho(y)$ which have a deep number-theoretical
origin. The behavior of $\rho(y)$ at small $y$ determines
the dependence of ${\cal D}$ and ${\cal C}$
on the pumping intensity $I\propto (\gamma\tau_{\varphi})^{-1}$.

For an infinite observation time ${\cal T}\rightarrow\infty$ the  function
$\rho(y)$ is {\it discontinuous} at any commensurate point $\alpha=P/Q$. 
For instance at $\alpha=1/2$ we have $\rho_{d}(y)=(4\pi)^{-1}\,\delta(y)$
and $\rho_{c}(y)=0$ at $y\ll 1$. However at any $\alpha\rightarrow
1/2$ the {\it smooth}
distribution $\rho_{d}(y)=\rho_{c}(y)$ does not show any peak at $y=0$.
This means that at a strong pumping the diffuson part of the variance of
conductance fluctuations drops dramatically if one goes away from
$\alpha=1/2$ and simultaneously the reduction factor moves towards the
universal value. This variation is discontinuous at ${\cal
T}\rightarrow\infty$ but at a finite observation time ${\cal T}$ (finite width
$\delta\sim 1/{\cal T}$ of the harmonic component) 
it happens at a scale
$\delta\alpha=\delta/\omega$.
This is illustrated by Fig.4 obtained by the numerical evaluation of  
Eqs.(11),(12) at zero temperature.
\begin{figure}[tbp]
\centerline{\epsfysize=5.2cm \epsffile{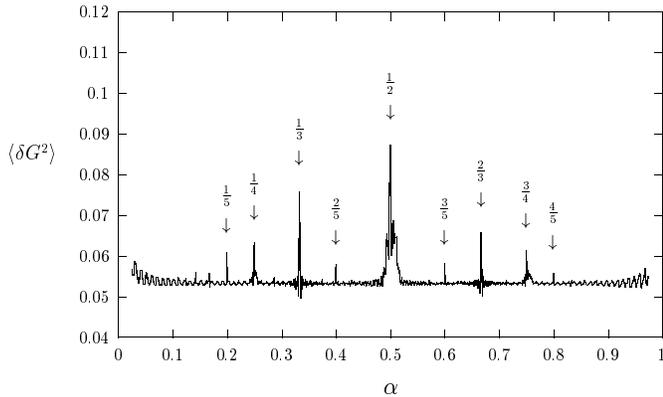}}
\caption{The total variance of conductance fluctuations (in units of the
ensemble-average dc conductance) as a function of
$\alpha$ at $\gamma\tau_{\phi}=0.05$ and
$1/{\cal T}\omega=0.01$. The width of peaks is of the order of $1/{\cal T}\omega$.}
\end{figure}

In conclusion, we have derived expressions for the diffuson and the
cooperon contributions to the variance of conductance fluctuations
in an open quantum dot subject to ac pumping. We focus at the difference
between the white--noise and  almost periodic ac pumping. 
We show that in the case of an almost periodic ac field the
pumping-induced dephasing is effectively switched off inside the `no-dephasing'
time windows. That
is why the periodic ac field is much less
effective in the suppression of conductance  
fluctuations than the noise-like ac field. 
In contrast to the ac noise, the periodic ac field may lead to the deviation of the
reduction factor $r$ (in the presence of a strong phase-breaking
mechanism) from the
universal value of 2. The dependence of the variance of conductance
fluctuations on
the pumping
intensity is studied for the harmonic pumping and the
quasi-periodic pumping with two harmonic components. In the latter case
we predict the new effect of {\it enchancement} of conductance
fluctuations
if two frequencies are commensurate.

We are grateful to I.L.Aleiner and I.V.Lerner for a clarifying discussion
and especially
to
E.Kanzieper and V.I.Yudson for a fruitful exchange of ideas during
collaboration on a similar problem.

\end{multicols}


\begin{references}
\bibitem{AAV} In the experimental geometry shown in Fig.2 there is no
restriction connected with the relationship between the pumping frequency
$\omega$ and the
Thouless energy $E_{c}$. However, if pumping is implemented by changing
of the dot shape or by the longitudinal time-dependent electric
field $E(t)$, the boundary conditions set the constraint on the
applicability of
Eqs.(\ref{dif1}-\ref{gamc}). They are valid only for $\omega\gg E_{c}$ \cite{YKK}.
For $\omega\ll E_{c}$ one can derive $\Gamma_{d}(\eta,\xi)-\gamma=C
(E(\xi+\eta/2)-E(\xi-\eta/2))^2$ and
$\Gamma_{c}(t,\xi)-\gamma=(C/4)
(E(t+\xi/2)-E(t-\xi/2))^2$ with $C\sim L^{2}/E_{c}$ \cite{VAA,YKK}. This leads only
to the re-definition of the parameter $\tau_{\varphi}^{-1}$ in Eq.(\ref{phi}):
$\tau_{\varphi}^{-1}=4C \overline{E^2}$.
\end{references}
\end{document}